\documentclass[11pt, oneside]{article}   	
\usepackage{geometry}                		
\usepackage{graphicx}				
\usepackage{tikz}

\usepackage{titling}
\usepackage{float}

\usepackage[pdftex,colorlinks=true,linkcolor=blue,citecolor=blue,urlcolor=black]{hyperref}						
						
\usepackage{thmtools,thm-restate}								
\usepackage{amssymb}
\usepackage{amsmath}
\usepackage{amsthm}
\usepackage{braket}
\usepackage{authblk}

\newtheorem{theorem}{Theorem}

\newtheorem{corollary}{Corollary}

\DeclareMathOperator{\Tr}{Tr}
\DeclareMathOperator{\polylog}{polylog}
\DeclareMathOperator{\E}{\mathbb{E}}
\DeclareMathOperator{\real}{Re}
\DeclareMathOperator{\poly}{poly}

\newcommand{\bqp}{{BQP}}
\newcommand{\ketbra}[2]{\left| #1 \rangle\langle #2\right|}

\title{The Quantum Complexity of Computing Schatten $p$-norms}

\author[]{Chris Cade\footnote{\nolinkurl{chris.cade@bristol.ac.uk}} }
\author[]{Ashley Montanaro}
\affil[]{School of Mathematics, University of Bristol, UK}

\begin{document}
\maketitle

\begin{abstract}
We consider the quantum complexity of computing Schatten $p$-norms and related quantities, and find that the problem of estimating these quantities is closely related to the one clean qubit model of computation. We show that the problem of approximating $\Tr(|A|^p)$ for a log-local $n$-qubit Hamiltonian $A$ and $p=\poly(n)$, up to a suitable level of accuracy, is contained in DQC1; and that approximating this quantity up to a somewhat higher level of accuracy is DQC1-hard. In some cases the level of accuracy achieved by the quantum algorithm is substantially better than a natural classical algorithm for the problem. The same problem can be solved for arbitrary sparse matrices in BQP. One application of the algorithm is the approximate computation of the energy of a graph.
\end{abstract}

\section{Introduction}
It is widely believed that quantum computers will be capable of solving certain computational problems more efficiently than any classical computer. However, the exact characterisation of the class of problems that allow for a quantum speedup is the subject of ongoing research. In complexity theory, this class is known as \bqp~\cite{watrous2009quantum} -- the set of languages efficiently decidable by a uniform family of polynomial-size quantum circuits with bounded error. A useful way to understand and identify the types of problems that are efficiently solvable by a quantum computer, but unlikely to be efficiently solvable by a classical computer, is to find problems that are complete for \bqp\footnote{We note that what we are really referring to here are PromiseBQP-complete problems, since there are in fact no known BQP-complete problems. For a detailed discussion on this point see \cite{janzing2007simple, goldreich2006promise}.}; that is, problems that can be solved by a polynomial-time quantum computer, and that any other problem in \bqp~can be reduced to. Intuitively, these are the very hardest problems in \bqp. 

Several \bqp-complete problems are known, including approximating the Jones polynomial \cite{aharonov2006polynomial}, estimating quadratically signed weight enumerators (QSWEs)\cite{knill2001quantum}, and estimating diagonal entries of powers of sparse matrices \cite{janzing2007simple}. The latter problem is particularly interesting, since it is a relatively natural problem that is not obviously `quantum' in nature.

Knill and Laflamme \cite{knill2001quantum} showed that a more constrained version of the QSWE problem is efficiently solvable in the one clean qubit model of computation -- an apparently non-universal model of quantum computation that is weaker than full quantum computation, but that can seemingly solve some problems more efficiently than a classical computer \cite{shepherd2006computation}. Understanding the power of such intermediate classes of computation could shed light on the types of problems that are efficiently solvable by a fully universal quantum computer. 

Here we consider the computational complexity of estimating Schatten $p$-norms of matrices. We find that for certain values of $p$ and certain families of matrices, this problem is closely related to the one clean qubit model of computation: depending on the accuracy of the estimation, the problem can be efficiently solved in the one clean qubit model, or is hard for this model of computation. We also consider similar quantities related to the spectra of matrices, such as the so-called ``energy'' of graphs~\cite{li2012graph, gutman2001energy}, and provide quantum algorithms for estimating them that are more efficient than any known classical algorithms.

\subsection{The One Clean Qubit Model of Computation}\label{sec:DQC1}
The one clean qubit model of quantum computation initially arose as an idealised model for computation on highly mixed initial states, such as those that appear in NMR implementations \cite{knill1998power}. In this model, we are given a quantum state consisting of a single `clean' qubit in the pure state $\ket{0}$, and $n$ qubits in the maximally mixed state. This can be represented by the density matrix
\[
\rho = \ketbra{0}{0} \otimes \frac{I_{2^n}}{2^n}. 
\]
We then apply an arbitrary polynomial-sized quantum circuit to $\rho$, and measure the first qubit in the computational basis. Following \cite{knill1998power}, we will refer to the class of problems that can be solved in polynomial time using this model of computation as DQC1 -- deterministic quantum computation with a single clean qubit. 

The canonical problem that can be solved in this model is that of estimating the normalised trace of a $2^n\times 2^n$ unitary matrix $U$ corresponding to a polynomial-size quantum circuit. This is achieved by applying a controlled version of $U$ to $\rho$, where the clean qubit is used as the control qubit and is put into the state $(\ket{0}+\ket{1})/\sqrt{2}$ using a Hadamard gate. More precisely, we apply the controlled-$U$ operator to the state 
\[
\rho' = \frac{1}{2}(\ket{0} + \ket{1})(\bra{0} + \bra{1}) \otimes \frac{I_{2^n}}{2^n}
\]
and then apply a Hadamard gate to the first qubit, before measuring it. The probability of measuring zero is $\frac{1}{2} + \frac{1}{2}\frac{\real(\Tr(U))}{2^n}$, which can be estimated up to accuracy $\epsilon$ by repeating the procedure $O(1/\epsilon^2)$ times. The imaginary part of the trace of $U$ can be estimated similarly by starting with the first qubit in the state $\frac{1}{\sqrt{2}}(\ket{0} - i\ket{1})$. This problem has been shown to be complete for the class DQC1 \cite{shor2008estimating}.
\\
\\
More generally, we might consider the DQCk model of computation. That is, deterministic quantum computation with $k$ pure qubits. If $k = O(\log(n))$, then the DQCk model is equivalent to DQC1 \cite{shor2008estimating}. This result is important for us since the quantum circuit that we apply to the initial state may require a number of ancilla qubits initialised to $\ket{0}$ in order to correctly perform its computation. For example, if the quantum circuit implementing the unitary $U$ performs the phase estimation routine, then it will usually require an additional $O(\log n)$ clean qubits. In the context of estimating the trace of a unitary matrix, this result tells us that it is possible in DQC1 to compute the trace of a sub-matrix whose size is an inverse-polynomially large fraction of the size of the input matrix.

\subsubsection{DQC1-complete Problems}
Knill and Laflamme~\cite{knill1998power} showed that the problem of estimating a coefficient in the Pauli decomposition of a quantum circuit, up to polynomial accuracy, is complete for the class DQC1. In fact, the aforementioned problem of estimating the normalised trace of a quantum circuit is a special case of this problem \cite{shor2008estimating}. Shor and Jordan~\cite{shor2008estimating} added to the relatively short list of DQC1-complete problems by showing that the problem of estimating Jones polynomials is also complete for the class DQC1. 

These quantities appear to be hard to compute classically, and therefore the one clean qubit model of computation seems to be more powerful than classical computation. However, it is unlikely that DQC1 contains all of BQP \cite{shepherd2006computation}, and thus this model of computation appears to have a computational power that is somewhere in between BPP and BQP. Some evidence in this direction was recently provided by Morimae \cite{morimae2017hardness}, who built on earlier work (\cite{morimae2014hardness}) to show that the output distribution of the one clean qubit model is difficult to sample from classically up to constant total variation distance error, provided that some complexity theoretic conjectures hold. 

Here we show that the problem of computing Schatten $p$-norms of matrices is also closely related to the class DQC1.


\subsection{Schatten $p$-norms and Graph Energy}
Schatten $p$-norms are ubiquitous in Quantum Information theory (see, for example, \cite{schatten1, schatten2, schatten3}). This family of matrix norms includes the three most commonly used norms in quantum information theory: the Schatten 1-norm is more commonly called the trace norm, the Schatten 2-norm is also known as the Frobenius norm, and the Schatten $\infty$-norm is called the operator norm or spectral norm. Here we consider the normalised Schatten $p$-norm, defined as 
\[
\|A\|_p := \left( \frac{\sum_j |\lambda_j|^p}{2^n} \right)^{1/p}
\]
for a $2^n\times 2^n$ Hermitian matrix $A$, where the sum ranges over the eigenvalues of $A$. 

For instance, the Schatten 1-norm is the average of the absolute values of the eigenvalues of $A$,
\[
\|A\|_1 = \frac{\Tr(|A|)}{2^n} = \frac{\sum_j |\lambda_j|}{2^n}.
\]
If we consider the matrix $A$ to be the adjacency matrix of a graph, this quantity is known as the `Graph Energy', and has applications in chemistry, where it is related to the total electron energy of a class of organic molecules \cite{li2012graph, gutman2001energy}. More generally, quantities relating to the spectra of adjacency matrices are used throughout Graph Theory to reveal information about the graphs that they represent. In the present work, we consider some `global' properties of the spectra of matrices and graphs -- i.e. those of the form $\Tr(f(A))/2^n$, for some suitably chosen function $f$. The Schatten $p$-norms are examples of such quantities.

\subsection{Our results}
We study the complexity of approximately computing the Schatten $p$-norms of sparse matrices and relate this to quantum computation. We consider Hermitian matrices of size $2^n \times 2^n$, where at most $d=\poly(n)$ entries in each row are non-zero, and call such matrices $d$-sparse. One fairly natural class of sparse matrices that can be expressed concretely is the class of `$\log$-local' Hamiltonians. That is, $k$-local $n$-qubit Hamiltonians, with $k = O(\log n)$ - i.e. Hermitian matrices that can be written as a sum
\[
A = \sum_{j=1}^{m}A_j,
\]
for some $m$, where each $A_j$ is a Hermitian matrix that acts non-trivially on at most $k = O(\log n)$ qubits. We assume that we are given the individual matrices $A_j$ directly, that $\|A_j\| = O(\poly(n))$ for all $j$, and that $m = \poly(n)$.

\begin{theorem}\label{theo:contained}
Let $A$ be a sparse Hermitian matrix on $n$ qubits, and let $p, 1/\epsilon = O(\poly(n))$. Then the problem of estimating $\frac{\Tr(|A|^p)}{2^n}$ up to additive accuracy $\epsilon \|A\|^p$ is contained in BQP. If the matrix $A$ is $\log$-local, then this problem is also contained in DQC1.
\end{theorem}

\begin{theorem}\label{theo:hard}
Let $A$ be a $\log$-local Hermitian matrix on $n$ qubits. Then the problem of estimating $\frac{\Tr(|A|^p)}{2^n}$ up to additive accuracy $\epsilon \left(\frac{\|A\|}{2}\right)^p$ for arbitrary $p, 1/\epsilon = O(\poly(n))$ is hard for the class DQC1. 
\end{theorem}

The BQP case of Theorem \ref{theo:contained} follows from the result of Janzing and Wocjan \cite{janzing2006bqp}, who gave a BQP algorithm for estimating diagonal entries of $f(A)$, for a sparse matrix $A$ and an appropriate function $f$ which can be taken to be $f(x) = |x|^p$.

We therefore see that the problem of computing Schatten $p$-norms for $p=O(\poly(n))$ is closely related to the one clean qubit model of computation. By contrast, for different values of $p$ the problem is related to other classes of computation. For instance, $\|A\|_\infty$ is the operator norm of $A$, and the problem of computing it approximately is QMA-complete\footnote{For a definition of the class QMA, see \cite{watrous2009quantum}.}, even for $2$-local Hamiltonians. To see this, suppose we have some upper bound $\Delta = O(\poly(n))$ on the largest eigenvalue of a $2$-local $n$-qubit Hamiltonian $A$. Define the matrix $B := \Delta I_{2^n} - A$. Then the largest eigenvalue of $B$ (in absolute value) corresponds to the smallest eigenvalue of $A$. Hence, if we can compute the smallest eigenvalue of $A$, then we can compute $\|B\|$, and vice versa. Since the problem of estimating the smallest eigenvalue of a $k$-local Hamiltonian is QMA-complete for $k \geq 2$ \cite{kempe2006complexity}, this implies QMA-completeness of the problem of estimating the operator norm of a $2$-local Hamiltonian. 

Note that the required accuracies of the estimates in Theorems \ref{theo:contained} and \ref{theo:hard} differ by a factor of $1/2^p$. Unfortunately, we were unable to reconcile this difference, and therefore we did not find a variant of the problem that is \emph{complete} for DQC1. 
\\
\\
Theorem \ref{theo:contained} gives us the following corollary:
\begin{corollary}
Let $A$ be a $\log$-local matrix corresponding to the adjacency matrix of a $2^n$-vertex graph $G$, and let $p, 1/\epsilon = O(\poly(n))$. The normalised Graph Energy of $G$, $\Tr(|A|)/2^n$, can be estimated up to additive accuracy $\epsilon \|A\|$ in DQC1.
\end{corollary}

In proving Theorem \ref{theo:contained}, we also show that there exists a polynomial-time quantum algorithm (in DQC1) for estimating $\Tr(A^p)/2^n$ up to error $\epsilon \|A\|^p$ for $1/\epsilon, p \in O(\poly(n))$. This is useful in the context of graph theory because it allows for an estimation of the expected number of closed walks that start from each vertex in a $2^n$-vertex graph. To obtain these algorithms, we prove a more general result: 

\begin{restatable}{lemma}{lemmain}
\label{lem:DQC1_containment}
For a log-local Hamiltonian $A$, and any $\log$-space polynomial-time computable function $f : I \rightarrow [-1,1]$ (where $I$ contains the spectrum of $A$) that is Lipschitz continuous with constant $K$ (i.e. $|f(x)-f(y)| \leq K|x-y|$ for all $x,y \in I$), there exists a DQC1 algorithm to estimate $\Tr(f(A))/2^n = \sum_j f(\lambda_j)/2^n$ up to additive accuracy $\epsilon (K + 1)$, where $\lambda_j$ denote the eigenvalues of $A$, and $\epsilon = \Omega(1/\poly(n))$. 
\end{restatable}

~\\
Often, one is interested in calculating the properties of general sparse matrices. We note that it is easy to give a quantum algorithm for estimating the above quantities for sparse matrices by making use of a result of Janzing and Wocjan \cite{janzing2007simple, janzing2006bqp}, who give a BQP algorithm for estimating the diagonal entries of $f(A)$, for some function $f$ that satisfies certain continuity constraints, but this comes at the expense of moving to the class BQP. 

\subsubsection{Estimating $\|A\|_p$}\label{sec:schatten_error}
Given a $\log$-local $n$-qubit Hamiltonian A, the algorithm of section \ref{sec:containment} outputs
\[
\Tr(|A|^p)/2^n \pm \epsilon\|A\|^p.
\]
By taking the $p$th root, we obtain an estimate of $\|A\|_p$ of the form
\begin{eqnarray*}
\left(\frac{\Tr(|A|^p)}{2^n} \pm \epsilon\|A\|^p\right)^{1/p} = \|A\|_p \left(1 + \frac{2^n\epsilon\|A\|^p}{\Tr(|A|^p)}\right)^{1/p}.
\end{eqnarray*}
The error will be small when $\Tr(|A|^p)$ takes a value close to its maximum of $2^n\|A\|^p$. In the best case, the relative error is close to 
\[
(1 + \epsilon)^{1/p}.
\]
This suggests that in these `good' cases, our algorithm can estimate $\|A\|_p$ up to a reasonable additive error in polynomial time. 
\\
\\
On the other hand, we can always bound 
\[
\frac{2^n\|A\|^p}{\Tr(|A|^p)} \leq \frac{2^n\|A\|^p}{2^n|\lambda_{\min}|^p} = \kappa(A)^p,
\]
where $\lambda_{\min}$ is the minimal eigenvalue of $A$ in absolute value, and $\kappa(A) = \|A\|\|A^{-1}\|$ is the condition number of $A$. In this case the relative error is at most
\[
\left(1 + \epsilon \kappa(A)^p \right)^{1/p}.
\]
Since we consider $p=\poly(n)$, the algorithm allows us to achieve relative error close to $\kappa(A)$ by taking $\epsilon = 1 - 1/\kappa(A)^p \approx 1$. Alternatively, we could achieve relative error $(1 + \delta)$ for some $\delta = O(1/\poly(n))$ by setting $\epsilon = ((1+\delta)^p-1)/\kappa(A)^p$. In this case, we sacrifice the run-time of the algorithm in order to improve the accuracy.

\subsection{Relation to Previous Work}
Our techniques are similar to those used in \cite{janzing2007simple} and \cite{harrow2009quantum}. In particular, we use the same combination of Hamiltonian simulation and phase estimation for estimating and manipulating the eigenvalues of a Hermitian matrix. To show DQC1-hardness, we use techniques from the Hamiltonian complexity literature, and in particular ideas due to Kitaev et al.~\cite{kitaev2002classical, kempe2006complexity}. 

By using a previous result of Janzing and Wocjan~\cite{janzing2007simple}, we can obtain a BQP algorithm for estimating $\Tr(A^p)/2^n$ for general sparse matrices; however, it is not clear how to implement this algorithm in DQC1, since it uses $O(n)$ ancilla qubits for the Hamiltonian simulation step. In \cite{janzing2007simple}, the authors describe a polynomial-time quantum algorithm for estimating the diagonal entries of the matrix $A^p$ up to error $\epsilon \|A\|^p$, for $\epsilon = O(1/\poly(n))$, and show that this problem is in fact BQP-complete for sparse symmetric matrices. The problem remains BQP-complete even for matrices with only $0,\pm1$ entries.

\subsection{Comparison with Classical Algorithms}

We were not able to find any previous results in the literature regarding the complexity of estimating the above quantities for sparse matrices. In Section \ref{sec:classical}, we give a classical algorithm for estimating the normalised trace of a sparse matrix raised to some power, and prove some bounds on the accuracy that this algorithm can achieve. 

We find that for some types of matrix, the value $\Tr(A^p)/2^n$ can be estimated efficiently classically, and for others, a quantum algorithm appears to have some advantage over a classical one. In general, we show the following:

\begin{theorem}
Given a $2^n \times 2^n$, $d$-sparse matrix $A$, there exists a classical algorithm to estimate $\Tr(A^p)/2^n$ up to accuracy $\epsilon d^p \|A\|_{\max}^p$ in time that is polynomial in $n, p $ and $1/\epsilon$, where $\epsilon = O(1/\poly(n))$ and $\|A\|_{\max}$ is used to denote the maximum absolute size of an entry in $A$.
\end{theorem}

Therefore, in the cases where $\|A\| \ll d \|A\|_{\max}$, we can get an advantage by making use of the algorithm of Theorem \ref{theo:contained}. We find that for certain classes of random graph (namely power-law graphs), the BQP algorithm for computing $\Tr(A^p)/2^n$ obtains a quadratic improvement in accuracy over the corresponding classical algorithm.

For $\log$-local Hamiltonians and constant $p$, there exists an efficient \emph{exact} classical algorithm for computing $\Tr(A^p)$. By using conventional matrix multiplication, it is possible to calculate the value of $A^p$ by multiplying the individual matrices $A_j$. This can be seen from the expression for $\Tr(A^p)$:
\[
\Tr(A^p) = \sum_{j_1,j_2,...,j_p} \Tr(H_{j_1}H_{j_2}\cdots H_{j_p}) ,
\]
where each index $j_i$ ranges from $1$ to $m$. Every $H_{j_i}$ is $k$-local, and the complexity of multiplying a $k$-local matrix by an $l$-local matrix is $O(2^{3(k+l)})$ (using a naive algorithm), and results in a $(k+l)$-local matrix. If we perform the matrix multiplications from left to right, then, for each term in the sum, the first multiplication will take time $O(2^{3(2k)})$, the second $O(2^{3(3k)})$, and so on, until the final multiplication takes time $O(2^{3(pk)})$. There will be $p-1$ of these multiplications performed in total, with each taking at most $O(2^{3\cdot pk})$ time, and hence the trace of $A_{j_1}A_{j_2}\cdots A_{j_p}$ can be calculated in $O(2^{3\cdot pk})$ steps. There are $m^p$ terms in the sum, and therefore the complexity of the entire computation is $O(m^p 2^{3\cdot pk})$. 

If we take $k = O(\log n)$ (i.e. take $A$ to be a $\log$-local Hamiltonian), the time complexity is $m^p n^{O(p)}$. For $p=O(1)$, this time complexity is polynomial and the output of this algorithm is better than the corresponding quantum algorithm, as it computes the desired value exactly. 

Note that the problem of computing $\Tr(|A|^p)$ appears to be substantially harder classically for odd $p$, since it cannot be found by simply computing powers of a matrix, and instead requires more knowledge about the eigenvalues of $A$. 

\subsection{Organisation}
We begin by providing a proof of Theorem \ref{theo:hard} in Section \ref{sec:hard}. Then section \ref{sec:containment} provides a proof of Theorem \ref{theo:contained}, by describing an algorithm in the one clean qubit model that can estimate $\Tr(f(A))/2^n$ for a $2^n\times 2^n$ $\log$-local matrix $A$ and an appropriately continuous function $f$. Following this, Section \ref{sec:classical} describes a classical algorithm for estimating $\Tr(A^p)$ and Section \ref{sec:comparison} compares the performance of this algorithm with its quantum counterpart for random sparse graphs. Appendices \ref{app:hamiltonians} and \ref{app:lipschitz} contain some calculations that are helpful in proving Theorem \ref{theo:contained}.


\section{Estimating $\Tr(|A|^p)/2^n$ is DQC1-hard}\label{sec:hard}
Here we show that the problem of estimating $\Tr(|A|^p)/2^n$ for a $2^n \times 2^n$ $\log$-local Hamiltonian $A$ up to a given accuracy is hard for the class DQC1. More precisely, we assume that we have access to an algorithm that can estimate $\Tr(|A|^p)/2^n$ up to accuracy $\epsilon \left( \frac{\|A\|}{2}\right)^p$, for $\epsilon = O(1/\poly(n))$ and $p = \poly(n)$, and show that this implies that we can solve any problem contained in DQC1. 

To do this we show that, given as input a real unitary $U$ (implemented by some polynomial-sized quantum circuit acting on $n$ qubits), it is possible to construct a $\log$-local Hamiltonian $A$ such that $\Tr(|A|^p)/2^n = \Tr(U)/2^n$, for some $p = \poly(n)$. Furthermore, we show that an estimation accuracy of $\epsilon\left( \frac{\|A\|}{2}\right)^p$ is sufficient to provide an estimate of $\Tr(U)/2^n$ up to accuracy $1/\poly(n)$. This problem is complete for the class DQC1 \cite{shor2008estimating}, which implies that the problem of estimating $\Tr(|A|^p)/2^n$ up to the stated accuracy is DQC1-hard.
\\
\\
The construction is based on ideas from Hamiltonian complexity, and in particular Kitaev's clock construction for the local Hamiltonian problem \cite{aharonov2002quantum}. We assume that we have a decomposition $U = U_{M-1}...U_1U_0$ of the circuit into $M$ elementary gates. Since $U$ is described by a polynomial-sized circuit, we have $M = \poly(n)$. We add $\lceil \log M\rceil$ additional qubits to act as a `clock' register, which is used to control the application of the individual unitaries, and define a unitary operator
\[
W := \sum_{l=0}^{M-1} \ketbra{l+1}{l} \otimes U_l,
\]
where addition is taken to be modulo $M$. It is straightforward to check that
\[
W^M = \sum_{l=0}^{M-1} \ketbra{l}{l} \otimes U_{l+M}...U_{l+2}U_{l+1}.
\]
Then we have 
\begin{eqnarray*}
\Tr(W^M) &=& \sum_{l=0}^{M-1} \Tr(\ketbra{l}{l}) \cdot \Tr(U_{l+M}...U_{l+2}U_{l+1}) \\
&=& \sum_{l=0}^{M-1} \Tr(U_M...U_2U_1) \\
&=& M\Tr(U),
\end{eqnarray*}
where the second step follows from invariance of the trace under cyclic permutations.

$W$ is $\log$-local with $m=\poly(n)$ terms, since each clock operator $\ket{l+1}\bra{l}$ acts on $\lceil \log M\rceil$ qubits, and each of the unitaries $U_l$ act on at most $O(1)$ qubits each. Define the Hermitian matrix
\[
A := \frac{1}{2}(W + W^\dag),
\]
Then the trace of $A^M$ gives the real part of the trace of $\frac{W^M}{2^M}$, since $A^M$ equals $1/2^M(W^M + W^{\dag^M})$ plus some other powers of $W$ and $W^\dag$ that are traceless, and therefore do not contribute to the trace of $A^M$. 

$W$ is a $2^{n+\lceil \log M \rceil} \times 2^{n+\lceil \log M \rceil}$ unitary matrix, and so we have $\|A\| \leq 1$. Thus, given the ability to estimate the normalised trace of $A^p$ up to accuracy $\left(\frac{\|A\|}{2}\right)^p \epsilon$, we can estimate the value of $\real[\Tr(U)]/2^n$ up to accuracy $1/\poly(n)$, which is the level of accuracy required for the class DQC1. To see this, we observe that, taking $p=M$ and assuming (without loss of generality) that $M$ is a power of 2,
\begin{eqnarray*}
\frac{\Tr(A^M)}{2^{n+ \log M }} \pm \frac{\epsilon}{2^M} &=& \frac{\real(\Tr(W^M))}{2^M2^{n+\log M}} \pm \frac{\epsilon}{2^M} \\
&=& \frac{M\real(\Tr(U))}{M2^M2^n} \pm \frac{\epsilon}{2^M}.
\end{eqnarray*}
Multiplying by $2^M$, we obtain
\[ 
\frac{\real(\Tr(U))}{2^n} \pm \epsilon,
\]
which is precisely the quantity that is DQC1-hard to compute. This is sufficient to show that the problem of estimating $\Tr(A^p)/2^n$ up to accuracy $\left(\frac{\|A\|}{2}\right)^p \epsilon$ for a $\log$-local $n$-qubit Hamiltonian is hard for the class DQC1. 
\\
\\
Note that we were not able to use standard techniques from the Hamiltonian complexity literature to make this construction work for $k$-local Hamiltonians with constant $k$ \cite{kempe2006complexity, kitaev2002classical}. These techniques involve the introduction of a larger clock space that is then acted upon by $k$-local Hamiltonians. A term is then added to the Hamiltonian to `penalise' invalid clock states and prevent them from contributing to the ground state energy. In our case, we care about the entire space on which the Hamiltonian acts and not just the subspace containing the valid clock states, and therefore the invalid clock states contribute to the trace of $A^M$ in a non-trivial way.


\section{Estimating $\Tr(|A|^p)/2^n$ is in DQC1}\label{sec:containment}

Here we show that the problem of estimating $\Tr(|A|^p)/2^n$ for a $\log$-local Hamiltonian $A$, up to reasonable error, is in DQC1. More precisely, we are given a $k$-local $n$-qubit Hamiltonian $A$, with $k = O(\log n)$; then the problem is to estimate $\Tr(|A|^p)/2^n$ up to error $\epsilon \|A\|^p$, for some integer $p = O(\poly(n))$ and accuracy $\epsilon = \Omega(1/\poly(n))$. We show that it is possible to construct a unitary $U$ such that the normalised trace of $U$ approximates the normalised trace of $|A|^p$. Moreover, we show that this construction can be performed in polynomial time (that is, the unitary $U$ takes $\poly(n,p,1/\epsilon)$ time to implement). In this way, we can use the DQC1 model to compute the normalised trace of the matrix $|A|^p$, hence showing that this problem is contained in DQC1. More generally, we show that it is possible to compute the value of $\Tr(f(A))/2^n$ for some function $f$, provided that it satisfies some continuity constraints. 

\lemmain*
Note that if the function $f$ does not map values in the interval $I$ to values in the interval $[-1,1]$ (e.g.  it might instead map $I \to \mathbb{R}$), then it suffices to compute the function $\bar{f}(x) := f(x)/f_{\max}$, where $f_{\max}$ is the supremum of $|f|$ on the interval $I$. Then, at the end of the computation, we can recover the original function by multiplying the output by $f_{\max}$. However, note that this will multiply the error of the algorithm by $f_{\max}$. Also note that the Lipschitz constant $K'$ of $f$ will be $f_{\max}K$, where $K$ is the Lipschitz constant of of $\bar{f}$. This gives the following corollary:
\begin{corollary}\label{corollary}
For a log-local Hamiltonian $A$, and any $\log$-space polynomial-time computable function $f : I \rightarrow \mathbb{R}$ (where $I$ contains the spectrum of $A$) that is Lipschitz continuous with constant $K'$ (i.e. $|f(x)-f(y)| \leq K'|x-y|$ for all $x,y \in I$), there exists a DQC1 algorithm to estimate $\Tr(f(A))/2^n = \sum_j f(\lambda_j)/2^n$ up to additive accuracy $\epsilon (K' + f_{\max})$, where $\lambda_j$ denote the eigenvalues of $A$, $\epsilon = \Omega(1/\poly(n))$, and $f_{\max}$ is the supremum of $|f|$ on the interval $I$. 
\end{corollary}

The proof of this lemma is split into roughly three parts. The first part (Section \ref{sec:circuit}) describes how the algorithm works. Following this, Section \ref{sec:errors} discusses the accuracy and failure probability of the algorithm, and finally, Section \ref{sec:clean_qubits} shows that the number of ancilla qubits required (and therefore the number of pure qubits needed) to implement the algorithm is at most $O(\log n)$.

\subsection{Constructing the Unitary}\label{sec:circuit}
We are given a $\log$-local Hamiltonian $A$ with eigenvectors $\ket{\psi_j}$ and corresponding eigenvalues $\lambda_j$. The basic idea is to construct a unitary $U$ whose eigenvalues correspond to the eigenvalues of $A$ in a useful way. In particular, we construct a polynomial-sized circuit whose associated unitary has eigenvalues $\lambda'_j$ such that $\lambda'_j = f'(\lambda_j)$, for some function $f'$ that depends on $f$.

The first step is to use Hamiltonian simulation to implement the unitary $e^{iA}$, which has eigenvalues $e^{i\lambda_j}$ for each eigenvector $\ket{\psi_j}$ of $A$. Section \ref{sec:hamiltonian_simulation} discusses the time complexity of this part of the circuit. Then the circuit performs the following sequence of operations, which we will describe in terms of their effects on an eigenvector $\ket{\psi_j}$ of $A$ and an arbitrary single qubit state of the form $\alpha\ket{0} + \beta\ket{1}$. We use $\ket{0}$ to denote an arbitrarily large ancilla register (with each qubit initialised to 0), and assume that both the phase estimation and Hamiltonian simulation parts of the circuit work perfectly.
\begin{enumerate}
\item Apply phase estimation on $e^{iA}$ with the input $\ket{\psi_j}$, to obtain an estimate of the eigenvalue $\lambda_j$: \[\ket{\psi_j}(\alpha\ket{0} + \beta\ket{1})\ket{0} \mapsto \ket{\psi_j}(\alpha\ket{0} + \beta\ket{1})\ket{\lambda_j}\] 

\item Perform controlled phase rotations, where the phase depends on a function $f$ of $\lambda_j$ contained in the 3rd register (for example, $f(x) = x^p$):
\[
\ket{\psi_j}(\alpha\ket{0} + \beta\ket{1})\ket{\lambda_j} \mapsto \ket{\psi_j}(\alpha e^{i \arccos(f(\lambda_j))}\ket{0} + \beta e^{-i \arccos(f(\lambda_j))}\ket{1})\ket{\lambda_j}
\]

\item Undo the phase estimation to uncompute the value in the 3rd register: 
\[
\mapsto \ket{\psi_j}(\alpha e^{i \arccos(f(\lambda_j))}\ket{0} + \beta e^{-i \arccos(f(\lambda_j))}\ket{1})\ket{0}
\]
\end{enumerate}
This gives us a unitary $U$ that performs the mapping
\[ 
\ket{\psi_j}(\alpha\ket{0}+\beta\ket{1})\ket{0} \mapsto (\alpha e^{+ i \arccos(f(\lambda_j))}\ket{\psi_j}\ket{0} + \beta e^{-i \arccos(f(\lambda_j))}\ket{\psi_j}\ket{1})\ket{0}
\]
for each eigenvector $\ket{\psi_j}$ of $A$. Therefore, for each eigenvalue $\lambda_j$ of $A$, $U$ has two corresponding eigenvalues $e^{\pm i \arccos(f(\lambda_j))}$.

By using the results described in Section \ref{sec:DQC1}, we can compute the trace of a sub-matrix of $U$ in the one clean qubit model, provided that the number of ancilla qubits used is $O(\log n)$ (we check that this is indeed the case at the end of this section). In particular, we compute the trace of $U'$, the sub-matrix of $U$ obtained by fixing the ancilla qubits (except the one explicitly mentioned above) to $\ket{0}$.
Then the trace of $U'$ is
\begin{eqnarray*}
\Tr(U') &=& \sum_j e^{\pm i \arccos (f(\lambda_j))} = \sum_j \cos(\pm \arccos(f(\lambda_j))) + i\sin(\pm \arccos(f(\lambda_j)))\\
&=& \sum_j 2\cos(\arccos(f(\lambda_j))) + i\sin(\arccos(f(\lambda_j))) - i\sin(\arccos(f(\lambda_j))) \\
&=& \sum_j 2f(\lambda_j) .
\end{eqnarray*}

\subsection{Error Analysis}\label{sec:errors}
Errors can arise in three places. Firstly, we will have some error in the Hamiltonian simulation part of the circuit. Secondly, there will be errors in estimating eigenvalues by using the phase estimation routine. And finally, there will be some error in the estimation of the normalised trace of $U$ from using the one clean qubit model. The analysis in this section is analogous to that of \cite{janzing2007simple}, since we use the same method for estimating an eigenvalue of $A$ via simulation of $e^{iA}$, but uses different methods to bound the errors introduced by phase estimation and Hamiltonian simulation.

\subsubsection{Error from Hamiltonian Simulation}

First we consider the error that arises from Hamiltonian simulation. We assume that the Hamiltonian simulation step implements a unitary $V$ that approximates $e^{iA}$ in the sense that $||V - e^{iA}|| \leq \delta$, so that the eigenvalues of $V$ and $e^{iA}$ can differ by at most $\delta$. For now, we will assume that the phase estimation routine works perfectly (i.e. introduces no error). Recall that this part of the circuit outputs an estimate for an eigenvalue of $A$ in the range $[-\pi, \pi)$. Denote by $\lambda_j$ and $\mu_j$ the output of the phase estimation routine when it is run using $e^{iA}$ and $V$, respectively. We have 
\[
\left| e^{i\lambda_j} - e^{i\mu_j} \right| \leq \delta
\]
by the bound on the error of the Hamiltonian simulation, where we can assume $|\mu_j - \lambda_j| \leq \pi$, by adding multiples of $2\pi$ to $\lambda_j$ if necessary. The left hand side can be written as
\begin{eqnarray*} 
\left| 1 - e^{i(\mu_j - \lambda_j)} \right| &=& \left| e^{i\frac{(\mu_j - \lambda_j) }{2}}\left(e^{-i\frac{(\mu_j - \lambda_j) }{2}} - e^{i\frac{(\mu_j - \lambda_j) }{2}}\right) \right| \\
&=& \left| e^{-i\frac{(\mu_j - \lambda_j) }{2}} - e^{i\frac{(\mu_j - \lambda_j) }{2}} \right| \\
&=& 2\left|\sin\left(\frac{\mu_j - \lambda_j}{2}\right)\right| \\
&=& 2\sin\left|\frac{\mu_j - \lambda_j}{2}\right| \qquad\qquad \text{(since $|\mu_j - \lambda_j| \leq 2\pi$)}.
\end{eqnarray*}
We will use the inequality 
\[
(2/\pi)\theta \leq \sin \theta 
\]
for $0 \leq \theta \leq \pi/2$. Therefore, we have that 
\[
(4/\pi)\frac{|\mu_j-\lambda_j|}{2} \leq 2\sin\left|\frac{\mu_j - \lambda_j}{2}\right| \leq \delta
\]
and hence
\[
|\mu_j-\lambda_j| \leq \pi\delta / 2.
\]
To see how this affects the accuracy of the algorithm, we consider the difference in the trace of $U'$ when using $V$ in place of $e^{iA}$. 
\begin{eqnarray*}
2\left| \sum_j f(\lambda_j) - \sum_j f(\mu_j)\right| &\leq& 2\sum_j \left| f(\lambda_j) - f(\mu_j)\right| \\
&\leq& 2\sum_j K\left| \lambda_j - \mu_j\right| \qquad \text{by the Lipschitz condition} \\
&\leq& 2\sum_j K\pi\delta/2 \\
&=& 2^nK\pi\delta.
\end{eqnarray*}
Choosing the simulation accuracy to be $\delta \leq \epsilon/(2\pi)$, this contributes an error term of $2^n \epsilon K/2$. Thus, we have
\begin{equation}\label{eq:error_after_HS}
2\left| \sum_j f(\lambda_j) - \sum_j f(\mu_j)\right| \leq 2^n \epsilon K/ 2.
\end{equation}

\subsubsection{Error from Phase Estimation}\label{sec:error_PE}
Here we consider the error that arises from using the phase estimation routine to estimate the eigenvalues $\mu_j$ of the unitary $V$ from the previous sub-section. The phase estimation routine requires the addition of $a$ ancilla qubits, which are used to control the application of powers of $V$ on an $n$-qubit register. More precisely, the $l$th ancilla qubit is used to control the application of the unitary $V^{2^l}$, so that we apply the controlled gate
\[
W_l := \ketbra{0}{0}_l \otimes I + \ketbra{1}{1}_l \otimes V^{2^l}
\]
where the subscript $l$ denotes that the projector acts on the $l$th ancilla/control qubit (and as the identity everywhere else). Let $W := W_1W_2\cdots W_a$. Then the phase estimation routine consists of applying Hadamard gates to all of the control qubits, applying $W$, and then applying the inverse quantum Fourier transform to the control qubits.

If we apply phase estimation to an eigenvector of $V$ with eigenvalue $e^{i2\pi\theta}$, and measure the control register, we obtain some output $x \in \{0,1,...,2^a-1\}$ such that 
\begin{equation}\label{eq:phase_est}
\Pr(|\theta - x/2^a| < \eta) > 1 - \varphi
\end{equation}
for $\varphi, \eta > 0$. To obtain this level of accuracy and probability of failure, it is sufficient \cite{nielsen2010quantum} to set 
\begin{equation}\label{eq:ancillas}
a = \lceil \log(1/\eta)\rceil + \lceil \log(2 + (1/(2\varphi)))\rceil.
\end{equation}
Let $\phi$ be defined as follows:
\[
\phi(x) := \begin{cases}
x2\pi / 2^a &\qquad \text{if } x \leq 2^{a-1} \\
x2\pi / 2^a - 2\pi &\qquad \text{otherwise}  \\
\end{cases}
\]
Then let $\phi(x_j)$ be our estimate of the eigenvalue $\mu_j$ corresponding to the eigenvector $\ket{\psi_j}$, which, by the definition of $\phi$ above, lies in the interval $[-\pi, \pi)$. By Equation (\ref{eq:phase_est}), if we apply phase estimation to an eigenvector $\ket{\psi_j}$ of $V$ with corresponding eigenvalue $e^{i\mu_j}$, and measure, we have 
\begin{equation}\label{eq:phase_promise}
\Pr(|\mu_j - \phi(x_j)| < 2\pi\eta) > 1 - \varphi
\end{equation}
where the extra factor of $2\pi$ results from rescaling the value of $x_j$ by $2\pi$. 
\\
\\
In our case, we do not measure the control register, and therefore we do not collapse the superposition over eigenvalues that phase estimation produces. Here we consider the effect that this has on the output of the algorithm, and simultaneously bound the error introduced by this part of the circuit. When phase estimation does not work perfectly, the algorithm consists of the following steps, implementing a unitary $\tilde{U}$:
\begin{enumerate}
\item Apply phase estimation on $V \approx e^{iA}$ with the input $\ket{\psi_j}$, to obtain a superposition over estimates $\phi(k)$ of the eigenvalue $\mu_j = 2\pi\theta_j$:
\[
\ket{\psi_j}(\alpha\ket{0} + \beta\ket{1})\ket{0} \mapsto \ket{\psi_j}(\alpha\ket{0} + \beta\ket{1})\sum_k \gamma_{k|j}\ket{\phi(k)} 
\]
where $\gamma_{k|j} = \frac{1}{N}\sum_a e^{2\pi ia(\theta_j - k/N)}$.

\item Perform controlled phase rotations:
\[
\ket{\psi_j}(\alpha\ket{0} + \beta\ket{1})\sum_k \gamma_{k|j}\ket{\phi(k)} \mapsto \ket{\psi_j}\sum_k \gamma_{k|j} (\alpha e^{i \arccos(f(\phi(k)))}\ket{0} + \beta e^{-i \arccos(f(\phi(k)))}\ket{1})\ket{\phi(k)}
\]

\item Undo the phase estimation to uncompute the value in the 3rd register. To undo phase estimation we: a) apply the QFT to the register containing the $\phi(k)$'s, b) apply controlled powers of the unitary $V^{\dag} \approx e^{-iA}$, and c) apply Hadamard gates to all qubits in the third register. 
\begin{enumerate}
\item Apply the QFT:
\[
\ket{\psi_j} \frac{1}{\sqrt{N}} \sum_k \gamma_{k|j}  (\alpha e^{i \arccos(f(\phi(k)))}\ket{0} + \beta e^{-i \arccos(f(\phi(k)))}\ket{1}) \sum_w e^{2\pi iwk / N}\ket{w}.
\]
\item Apply the controlled (on the third register) $V^\dag$ gates:
\[
\ket{\psi_j} \frac{1}{\sqrt{N}} \sum_k \gamma_{k|j} (\alpha e^{i \arccos(f(\phi(k)))}\ket{0} + \beta e^{-i \arccos(f(\phi(k)))}\ket{1})  \sum_w  e^{2\pi iwk / N} e^{-2\pi i\theta_j w} \ket{w}.
\]
\item Apply Hadamard gates to each of the ancilla qubits:
\[
\ket{\psi_j} \frac{1}{N} \sum_k \gamma_{k|j} (\alpha e^{i \arccos(f(\phi(k)))}\ket{0} + \beta e^{-i \arccos(f(\phi((k)))}\ket{1})  \sum_w \sum_x  e^{2\pi iwk / N} e^{-2\pi i\theta_j w} (-1)^{w\cdot x} \ket{x}.
\]
\end{enumerate}
\end{enumerate}
This means that $\tilde{U}$ performs the mapping
\[
\ket{\psi_j}(\alpha\ket{0} + \beta\ket{1})\ket{0} 
\]
\[
\mapsto \ket{\psi_j} \frac{1}{N} \sum_k \gamma_{k|j} (\alpha e^{i \arccos(f(\phi(k)))}\ket{0} + \beta e^{-i \arccos(f(\phi(k)))}\ket{1})  \sum_x \left(\sum_w  e^{2\pi iwk / N} e^{-2\pi i\theta_j w} (-1)^{w\cdot x}\right) \ket{x}
\]
for each eigenvector $\ket{\psi_j}$ of $V$.

Let $\{\ket{\psi_j}\ket{b}\ket{\phi}$, $b\in\{0,1\}\}$ be a basis for the tensor product of the three registers. By design, the only states that contribute to the trace of $U'$ are those of the form $\ket{\psi_j}\ket{b}\ket{0}$. Hence, we can consider the trace of $\tilde{U'}$ -- the submatrix of $\tilde{U}$ in which the third register is in the state $\ket{0}$) -- which is given by:
\begin{eqnarray*}
\Tr(\tilde{U'}) &=& \sum_{j} \bra{\psi_j}\bra{0} \left( \ket{\psi_j} \frac{1}{N} \sum_k \gamma_{k|j} \sum_w  e^{2\pi iwk / N} e^{-2\pi i\theta_j w} e^{i \arccos(f(\phi(k)))}\ket{0}    \right) \\
&+& \sum_{j} \bra{\psi_j}\bra{1} \left( \ket{\psi_j} \frac{1}{N} \sum_k \gamma_{k|j} \sum_w  e^{2\pi iwk / N} e^{-2\pi i\theta_j w} e^{-i \arccos(f(\phi(k)))}\ket{1}   \right) \\
&=& \frac{1}{N} \sum_{j,k} \gamma_{k|j} \sum_w  e^{2\pi iwk / N} e^{-2\pi i\theta_j w} \left(e^{i \arccos(f(\phi(k)))}+e^{-i \arccos(f(\phi(k)))}\right) \\
&=& \frac{1}{N} \sum_{j,k} \gamma_{k|j} 2f(\phi(k)) \sum_w  e^{2\pi iw(k / N -  \theta_j)}  \\ 
&=& 2 \sum_{j,k} \left|\gamma_{k|j}\right|^2 f(\phi(k))  \\
&=& 2 \sum_{k} f(\phi(k)) \sum_j \left|\gamma_{k|j}\right|^2 . 
\end{eqnarray*}
Suppose that $\theta_j = z_j/N$ for some $z_j$ -- that is, each $\theta_j$ can be represented precisely by an $n$-bit rational number $z_j/N$. Then $\gamma_{k|j} = \delta_{k,z_j}$, and so $\Tr(\tilde{U'}) = 2\sum_j f(\mu_j)$. This corresponds to the case in which phase estimation works perfectly; in reality, we will not be able to express all eigenvalues precisely as $n$-bit rational numbers. Instead, suppose that $\theta_j = \tilde{z}_j/N + \delta_j$, where $\tilde{z}_j/N$ is the closest $n$-bit approximation of $\theta_j$, and so $0 \leq \delta_j \leq 1/(2N)$. The difference between the trace in the two cases is given by
\begin{eqnarray*}
2 \left| \sum_{j} f(\mu_j) - \sum_{j} \sum_{k} \left|\gamma_{k|j}\right|^2f(\phi(k)) \right| &\leq& 2 \sum_{j} \left| f(\mu_j) - \sum_{k} \left|\gamma_{k|j}\right|^2f(\phi(k)) \right| \\
&=& 2 \sum_{j} \left| \sum_{k}\left|\gamma_{k|j}\right|^2(f(\mu_j) - f(\phi(k))) \right|  \\
&\leq& 2 \sum_{j} \sum_{k} \left|\gamma_{k|j}\right|^2 \left| f(\mu_j) - f(\phi(k)) \right| ,
\end{eqnarray*}
where the second step follows because $\sum_k \left|\gamma_{k|j}\right|^2 = 1$. The coefficient $\left|\gamma_{k|j}\right|^2$ is precisely the probability of measuring $\phi(k)$ on the ancilla register when the true eigenvalue is $\mu_j$. By the promises of phase estimation (Equation (\ref{eq:phase_promise})), with probability $ \leq \varphi$ we have $\left| \mu_j - \phi(k) \right| > 2\pi\eta$, in which case $\left| f(\mu_j) - f(\phi(k)) \right| \leq 2f_{\max}$; and with probability $\geq 1 - \varphi$ we have $\left| \mu_j - \phi(k) \right| \leq 2\pi\eta$, in which case $\left| f(\mu_j) - f(\phi(k)) \right| \leq 2\pi K\eta$. Hence, the error from this part of the circuit is bounded above by
\[
2 \left| \sum_{j} f(\mu_j) - \sum_{j} \sum_{k} \left|\gamma_{k|j}\right|^2f(\phi(k)) \right| \leq 4\sum_j (\pi K\eta + \varphi f_{\max}) = 2^{n+2} (\pi K\eta + \varphi f_{\max}).
\]
Choosing $\eta < \epsilon/(8\pi)$ and $\varphi < \epsilon/8$, and assuming that $f_{\max} \leq 1$ (as stated earlier), this becomes
\begin{equation}\label{eq:general_error2}
2 \left| \sum_{j} f(\mu_j) - \sum_{j} \sum_{k} \left|\gamma_{k|j}\right|^2f(\phi(k)) \right| \leq 2^{n} \frac{1}{2}\epsilon(K + 1).
\end{equation}
Now we consider how this contributes to the overall error. As before, let $\lambda_j$ denote the eigenvalues of $e^{iA}$. Then the error of the algorithm, taking into account both the Hamiltonian simulation and phase estimation steps, is 
\[
2 \left| \sum_{j} f(\lambda_j) - \sum_{j} \sum_{k} \left|\gamma_{k|j}\right|^2f(k) \right| \leq 2\left| \sum_j f(\lambda_j) - \sum_j f(\mu_j)\right| + 2\left| \sum_{j} f(\mu_j) - \sum_{j} \sum_{k} \left|\gamma_{k|j}\right|^2f(k) \right| 
\]
where the first term on the right corresponds to the error from the Hamiltonian simulation part of the circuit (i.e. the difference between the trace of the circuit when using $V$ instead of $e^{iA}$), and the second term corresponds to the error introduced by phase estimation. A bound on the first term is given by Equation (\ref{eq:error_after_HS}), and the second term is bounded via Equation (\ref{eq:general_error2}). Therefore, the difference in the trace of $U'$ in the case where Hamiltonian simulation and phase estimation both work perfectly, and when they do not, is bounded by
\begin{equation}\label{eq:HS_PS_error}
2 \left| \sum_{j} f(\lambda_j) - \sum_{j} \sum_{k} \left|\gamma_{k|j}\right|^2f(k) \right| \leq 2^n\epsilon(K + 1/2)
\end{equation}

\subsubsection{Error from estimating $\Tr(U')/2^n$ in the DQC1 model}
The one clean qubit model can estimate the normalised trace of a $2^n \times 2^n$ sub-matrix of a $2^{n + O(\log n)}\times 2^{n + O(\log n)}$ unitary matrix (implemented by a $\poly(n)$-sized circuit) up to accuracy $\zeta = \Omega(1/\poly(n))$. Therefore, using the one clean qubit model to estimate the trace of $U'$ will introduce an extra error term $\zeta$. Let $\widetilde{\Tr(U')}/2^n$ be the output from the one clean qubit algorithm. Then choosing $\zeta = \epsilon / 2$, and using the bound from Equation (\ref{eq:HS_PS_error}), we have
\begin{equation}\label{eq:final_error}
\left| \frac{2}{2^n}\sum_j f(\lambda_j) - \widetilde{\Tr(U')}/2^n \right| \leq \epsilon(K + 1).
\end{equation}
Hence, we can estimate $\frac{1}{2^n}\sum_j f(\lambda_j)$ in polynomial time with accuracy $\epsilon(K + 1)$ for any $\epsilon = \Omega(1/\poly(n))$.

\subsection{How many clean qubits are needed?}\label{sec:clean_qubits}
Here we consider how many clean qubits are required to implement the circuit described in Section \ref{sec:circuit} up to the desired accuracy. Any time the circuit uses ancilla qubits, these qubits will generally need to be initialised in the all-zeros state -- that is, they must be under our control, and be `clean'. As discussed in Section \ref{sec:DQC1}, we can use $O(\log n)$ clean qubits without changing the model of computation. In this section we argue that the implementation of the circuit described above requires no more than $O(\log n)$ ancilla qubits. 

The two main parts of the circuit are the phase estimation routine, and Hamiltonian simulation. The rest of the circuit consists of more basic operations that require only a constant number of ancilla qubits (provided that the function $f$ we choose is sufficiently easy to compute). In Section \ref{sec:errors}, we set the parameters for phase estimation $\theta$ and $\eta$ to be $\eta < \epsilon/8\pi$ and $\varphi < \epsilon/8$, where $\epsilon$ is inverse polynomial in $n$. Then by Equation (\ref{eq:ancillas}), the number of ancilla qubits required to implement the phase estimation part of the circuit is $O(\log n)$.

In order to implement the simulation of the Hamiltonian $A$, we can use techniques based on the Lie-Trotter product formula \cite{lloyd1996universal}. This requires no more than a constant number of ancilla qubits, and, since we assume that we are given the Hamiltonian directly as a set of $m$ individual Hamiltonians that each act on $O(\log n)$ qubits, there are no ancilla qubits required to `load' the input into the system, which would be the case if we considered the case where the input Hamiltonian is specified by an oracle (it is precisely for this reason that we define the problem in terms of a $\log$-local Hamiltonian rather than a sparse Hamiltonian). In our case, we can run a polynomial-time classical algorithm to compute the quantum circuit required to implement the unitary $e^{iA}$, given such a description of $A$. This is discussed more fully in the following section.

\subsection{Simulating log-local Hamiltonians}\label{sec:hamiltonian_simulation}
We are required to implement the unitary $e^{iA}$ for some $\log$-local Hamiltonian $A$. We are limited to using at most $O(\log n)$ ancilla qubits, which rules out the more advanced Hamiltonian simulation techniques that are based on quantum walks (e.g. \cite{berry2015hamiltonian,berry2016corrected}). Instead, we use the `vanilla' version of Hamiltonian simulation, which is based on the Lie-Trotter product formula \cite{lloyd1996universal}.

We are given a $\log$-local $n$-qubit Hamiltonian $A$, and wish to implement a unitary operator that approximates $e^{i A t}$ for some value of $t$, up to a specified accuracy $\delta$ (in the operator norm). That is, we want to construct, in classical polynomial time, a quantum circuit that implements a unitary operator $V$ such that
\[
\|V - e^{iAt}\| \leq \delta.
\]
In Appendix \ref{app:hamiltonians} we check that the standard techniques, which are usually presented for $O(1)$-local Hamiltonians, indeed work for $\log$-local Hamiltonians, and confirm that we can simulate $e^{iAt}$ up to accuracy $\delta$ in time 
\[
O(\poly(m,n,\tau,1/\delta)),
\]
where $\tau = t\|A\|$, using a circuit that can be computed by a polynomial-time classical algorithm. The time complexity could be improved by the use of more complicated simulation techniques \cite{berry2007efficient}, but we do not consider this here. 
\\
\\
In the circuit described in Section \ref{sec:circuit}, we set $t=1$, and require that $\delta = O(1/\poly(n))$. Thus, the time taken to implement the Hamiltonian simulation part of the circuit will be $O(\poly(n))$. 

\subsection{Proof that estimating $\Tr(|A|^p)/2^n$ is in DQC1}\label{sec:containment_schatten}
The proof of Theorem \ref{theo:contained}, which states that the problem of estimating $\Tr(|A|^p)/2^n$ up to error $\epsilon \|A\|^p$ is in DQC1 for $p, 1/\epsilon = \poly(n)$, follows almost immediately from Lemma \ref{lem:DQC1_containment}. The same proof also applies to the problem of estimating $\Tr(A^p)/2^n$. 

It is straightforward to check that, on the interval $[-b,b]$, both $f(x) = x^p$ and $f(x) = |x|^p$ are Lipschitz continuous with Lipschitz constant $K = pb^{p-1}$ (see Appendix \ref{app:lipschitz}). Furthermore, we have $f_{\max} = b^p$ for both functions. In our case we can take $b = \|A\|$ since $f$ is a function of the eigenvalues of $A$. We can then apply Corollary \ref{corollary} to $f$. 

Putting these values into Corollary \ref{corollary}, and replacing $\epsilon$ with $\frac{\epsilon}{p/\|A\| + 1}$, we obtain an estimate of $\frac{\Tr(|A|^p)}{2^n}$ up to accuracy $\epsilon \|A\|^p$. Furthermore, this estimate can be obtained in DQC1 in time that is polynomial in $n$ and inverse polynomial in $\epsilon$.

\section{Classical Algorithms}\label{sec:classical}
We next describe a classical algorithm for diagonal entry estimation, which is the problem of estimating an entry on the diagonal of the matrix $A^p$, up to reasonable error. Given the ability to estimate the diagonal entries of a matrix, we are able to estimate the normalised trace of the matrix.  
\\
\\
We first present an algorithm for the special case where $A$ contains only $0,1$ entries, and then discuss how it can be extended to work for arbitrary real matrices. In the first case, the matrix $A$ defines an unweighted, undirected graph with $N$ vertices. The value of $(A^p)_{jj}$ is equivalent to the number of distinct walks (i.e. traversals around the graph that may traverse any edge more than once, or not at all) of length $p$ starting and ending at vertex $j$. 

We begin by observing that $(A^p)_{jj}$ can be re-interpreted as the total number of walks leaving $j$ of length $p$ multiplied by the probability that such a walk ends at vertex $j$. We can obtain an estimate of the latter by performing a number of random walks of length $p$, beginning at vertex $j$, and counting how many of them return to vertex $j$ on the final step. 

In order to obtain an estimate of the total number of walks of length $p$ leaving a given vertex, we can do the following: given an upper bound $d$ on the degree of the graph, we generate a number of sequences of $p$ integers chosen independently and uniformly at random from the range $[0, d]$. Any given sequence provides a `candidate' walk of length $p$ on the graph, which may or may not be `realisable' on the graph defined by $A$. Given a candidate walk of the form $(n_0, n_1, ... , n_p)$, we test whether or not it is realisable by starting a walk at vertex $j$, and then moving to the $n_0$th neighbour of $j$. We then move to the $n_1$th neighbour of that vertex, and so on. If, at any step $i$ of the walk, a vertex does not have a neighbour $n_i$, we terminate the process and conclude that the candidate is not realisable. 

If we tried all $d^p$ possible candidate walks from vertex $j$, then by counting the number of successes we would know the exact value of the number of walks of length $p$ that leave vertex $j$; however, this would require $O(d^p)$ walks to be performed. If instead we sample from the set of all possible walks by generating a number of sequences at random, we can obtain a close estimate of the true number of walks. 

Below is the full algorithm for diagonal entry estimation. We assume that we are given some bound $d$ on the degree of the graph, and that we wish to estimate $(A^p)_{jj}$. 
\begin{enumerate}
\item Estimate the total number of walks of length $p$ leaving vertex $j$:
\begin{enumerate}
	\item Define variables $X_i$ for $i \in [k]$, for some value of $k$ to be determined later.
	\item For $i = 1$ to $k$:
	\begin{enumerate}
		\item Generate a sequence $(n_0, n_1, ... , n_p)$, where each $n_l \in [d]$.
		\item Attempt to follow the walk defined by the sequence.
		\item If the walk was successful, set $X_i = 1$, otherwise set $X_i = 0$. 
	\end{enumerate}
	\item Then $\overline{X} = \frac{d^p}{k}(X_1 + X_2 + ... + X_k)$ provides an estimate of the total number of walks of length $p$ leaving vertex $j$. 
\end{enumerate}
\item Estimate the probability that a given walk returns to vertex $j$:
\begin{enumerate}
	\item Define variables $Y_i$ for $i \in [k']$, for some value of $k'$ to be determined later.
	\item For $i = 1$ to $k'$:
	\begin{enumerate}
		\item Perform a random walk of length $p$ starting at vertex $j$.
		\item If the walk returns to vertex $j$ (as its final step), then set $Y_i = 1$, otherwise set it to $0$.
	\end{enumerate}
	\item Then $\overline{Y} = \frac{1}{k'}(Y_1 + Y_2 + ... + Y_{k'})$ gives an estimate of the probability that a given walk returns to vertex $j$.
\end{enumerate}
\item Multiplying the two values together gives us our desired estimate: $(\tilde{A^p})_{jj} = \overline{X} \cdot \overline{Y}$.
\end{enumerate}
To analyse the accuracy of this estimation, we will look at the errors in the two estimates $\overline{X}$ and $\overline{Y}$.
\\
\\
In both steps, we are essentially aiming to estimate the probability of success of some Bernoulli process: in step 1 we aim to estimate the probability with which a randomly generated sequence of `moves' succeeds in generating a valid walk around the graph, and in step 2 we are estimating the probability that a given (valid) walk of length $p$ succeeds in returning to its starting vertex on the final step of the walk. In both cases, we can estimate the appropriate probability up any desired accuracy $\epsilon$ by choosing the number of samples ($k$ in step 1, and $k'$ in step 2) to be inverse polynomial in $\epsilon$. 

We use Hoeffding's inequality to bound the accuracy of both estimates. For step 1, we absorb the factor of $d^p$ into the random variables $X_i$, and use the general form of the bound:
\[
\Pr\left[ |\overline{X} - \E[\overline{X}] | \geq \epsilon d^p\right] \leq 2e^{-2\epsilon^2 k}.
\]
And for step 2, we have 
\[
\Pr\left[ |\overline{Y} - \E[\overline{Y}] | \geq \epsilon' \right] \leq 2e^{-2\epsilon'^2 k'}.
\]
Therefore, by choosing $k = \poly(1/\epsilon)$ and $k' = \poly(1/\epsilon')$, we can estimate $(A^p)_{jj}$ up to additive error that is at most $d^p (\epsilon + \epsilon' + \epsilon\epsilon) = d^p\delta$ for $\delta = 1/\poly(n)$, with a constant probability of failure.

\subsection{Extension to more general matrices}
The above algorithm works for matrices with $0,1$ entries by interpreting the input matrix as the adjacency matrix for an unweighted, undirected graph. More general (symmetric) matrices may be interpreted as undirected graphs with weighted edges. A similar interpretation of the value of $(A^p)_{jj}$ holds in these cases. We will begin by extending the algorithm to matrices with $-1, 0, +1$ entries. In this case, the value of $(A^p)_{jj}$ depends not only on the number of closed walks (i.e. those that return to their start vertex) leaving vertex $j$, but also on the `parity' of those walks. That is, $(A^p)_{jj}$ gives the total number of closed walks with even parity minus the number of closed walks with odd parity, where the parity of the walk is even if there are an even number of edges on the walk with a weight of $-1$, and odd otherwise. The value of $(A^p)_{jj}$ can then be computed as
\[
W_p \times [\Pr(\text{Walk returns to $j$ with even parity}) - \Pr(\text{Walk returns to $j$ with odd parity})]
\]
where $W_p$ is the total number of walks of length $p$ leaving vertex $j$, which can be estimated using the same approach as before. It is also straightforward to estimate the two probabilities using a similar method to the previous algorithm. In fact, we can combine the two cases and define a set of variables $Y_i$ for $i \in k'$, setting $Y_i = +1$ if the $i$th random walk returned to $j$ with even parity, $-1$ if it returned to $j$ with odd parity, and $0$ otherwise. Then the algorithm proceeds as before. 

Since each $Y_i$ can take values in the range $[-1,+1]$, the accuracy is changed, and we obtain a slightly different result from Hoeffding's inequality:
\[
\Pr\left[ |\overline{Y} - \E[\overline{Y}] | \geq \epsilon' \right] \leq 2e^{-\epsilon'^2 k'}. 
\]
By choosing $k$ and $k'$ as before (i.e. as inverse polynomials in $\epsilon$ and $\epsilon'$, respectively), we obtain the same accuracy of $\delta d^p$ with constant probability. 

It is interesting to note that in the case of $+1, 0, -1$ matrices, the accuracy of the estimation does not change (up to constant factors). The difference between the accuracies achieved by the classical and quantum algorithms therefore lies in the difference between the values of $\|A\|$ and $d$. 
\\
\\
Now we move to the more general case of an arbitrary sparse (symmetric) real matrix. In this case, the interpretation of $(A^p)_{jj}$ is a little more complicated. Let $\mathcal{C}^j_p$ be the set of all \emph{closed} walks of length $p$ leaving vertex $j$, and $E(\omega)$ be the set of edges that make up a given walk $\omega$. 

Then we have
\[
(A^p)_{jj} = \sum_{c \in \mathcal{C}^j_p} \prod_{e\in E(c)} \text{weight}(e).
\]
In order to estimate this quantity, we proceed similarly to the above two cases. 
Let us denote the set of all (not necessarily closed) walks of length $p$ originating at vertex $j$ by $\mathcal{W}^j_p$. Then we can re-write the above quantity as
\[
(A^p)_{jj} = W_p \E_{\omega\in\mathcal{W}^j_p}\left[\prod_{e\in E(\omega)} \text{weight}(e)\right],
\]
by using the same reasoning as before -- i.e. that the $j$th diagonal entry of $A^p$ is given by the total number of walks of length $p$ leaving vertex $j$ multiplied by the expected `weight' of each walk, where we assign a weight of $0$ if the walk does not return to vertex $j$. 

We can estimate the expectation on the right by sampling from the set of closed walks of length $p$ originating at vertex $j$. This can be done by performing random walks of length $p$ starting at vertex $j$, and recording the total weights of those walks that return to vertex $j$. This is easily incorporated into the existing algorithm: we set the variable $Y_i$ to 0 if the $i$th walk does not return to vertex $j$, and otherwise we set it to the total weight of the walk (i.e. the product over the weights of the edges of the walk). $W_p$ can be estimated as before, up to error $\epsilon d^p$. The error in estimating the expectation value depends upon the largest total weight of a closed walk in the graph. This is smaller than or equal to $\|A\|_{\max}^p$, where $\|A\|_{\max}$ is the maximum absolute size of an entry in $A$.

A bound on the accuracy of estimating the expectation value is once again given by Hoeffding's inequality:
\[
\Pr[|\overline{Y} - \E[\overline{Y}]| \geq \epsilon'\|A\|_{\max}^p] \leq 2e^{-2\epsilon'^2k'}.
\]
Multiplying the two estimates together, we obtain an estimate of $(A^p)_{jj}$ up to accuracy $\delta d^p\|A\|_{\max}^p$ with constant probability. 

\subsection{Estimating $\Tr(A^p)/N$ Classically}
We can use the classical version of diagonal entry estimation to estimate the normalised trace of a matrix. More precisely, we obtain the empirical mean of $(A^p)_{jj}$ over a sample of values of $j$ chosen uniformly at random. To see that the mean value of $(A^p)_{jj}$ for $j \in [N]$ does indeed give us the desired value, we observe that 
\[
\E_j[(A^p)_{jj}] = \frac{1}{N} \sum_{j = 0}^{N-1} (A^p)_{jj} = \frac{\Tr(A^p)}{N}.
\]
Let the output of the diagonal entry estimation algorithm be $(\widetilde{A^p})_{jj}$ (which is an estimate of $(A^p)_{jj}$ up to additive error $\delta d^p \|A\|_{\max}^p$). Then let $\overline{(\widetilde{A^p})_{jj}}$ be the mean value of the variable $(\widetilde{A^p})_{jj}$ after sampling $k$ times for randomly chosen values of $j$. The value of $(\widetilde{A^p})_{jj}$ is bounded in the interval $[-(d\|A\|_{\max})^p, (d\|A\|_{\max})^p]$. Then by Hoeffding's inequality:
\[
\Pr\left[ \left| \overline{(\widetilde{A^p})_{jj}} - \E[(\widetilde{A^p})_{jj}] \right| \geq \delta d^p \|A\|_{\max}^p \right] 
\leq 2 \exp \left( \frac{-\delta^2}{2}k\right).
\]
Thus, choosing $k$ to be inverse polynomial in $\delta$ allows us to obtain an estimate of $\E[(A^p)_{jj}] = \Tr(A^p)/N$ up to error $\delta d^p\|A\|_{\max}^p$. Note that for $0,1$ and $-1,0,+1$ matrices, $\|A\|_{\max} = 1$ and therefore the accuracy of the estimation in this case is just $\delta d^p$. 

We are now in a position to compare the performance of the quantum algorithm for trace estimation to the classical counterpart for various families of matrices.

\section{Quantum vs. Classical}\label{sec:comparison}
We compare the complexities of the (BQP) quantum and classical algorithms for computing $\Tr(A^p)$, for random $N\times N$ matrices. Recall that the quantum algorithm has an accuracy of $\epsilon \|A\|^p$, and that the classical algorithm has an accuracy of $\epsilon d^p$ in the $0,1$ and $-1,0,1$ cases, and an accuracy of $\epsilon d^p\|A\|_{\max}^p$ in the general case, where $p = \polylog(N)$. 

In the event that $\|A\| \ll d$, the quantum algorithm achieves an improvement in accuracy over the classical algorithm. However, since the quantum algorithm requires the matrix $A$ to be sparse, we must restrict our attention to only sparse matrices that have this property. Towards this end, we will begin by considering a general model for random graphs, and introduce some results that relate the degrees of the vertices of the graph to the eigenvalues of the adjacency matrix. Following this, we will consider how these results apply to sparse graphs.

\subsection{Random Graphs}

We consider a general model for unweighted random graphs (see e.g. \cite{chung2003spectra}), in which each vertex $v$ is associated with a weight $w_v$. Then a random graph $G$ is constructed by assigning an edge independently to each pair of vertices $(i,j)$ with probability $\frac{w_iw_j}{\sum_{i} w_i}$, such that the expected degree of vertex $v$ is given by $w_v$. Denote by $d$ the maximum expected degree, and by $\tilde{d}$ the value
\[
\tilde{d} := \frac{\sum_{i=1}^N w_i^2}{\sum_{i=1}^N w_i}.
\]
Then we have the following results from \cite{chung2003spectra}:
\begin{theorem}\label{theo:d_bounds}
If $\tilde{d} > \sqrt{d} \ln N$, then as $N\to \infty$ the largest eigenvalue of a random graph $G(w)$ is almost surely $(1+o(1))\tilde{d}$.
\end{theorem}
\begin{theorem}
If $\sqrt{d} > \tilde{d} \ln^2 N$, then as $N\to \infty$ the largest eigenvalue of a random graph $G(w)$ is almost surely $(1+o(1))\sqrt{d}$
\end{theorem}
Intuitively, $\|A\|$ is (asymptotically) the maximum of $\sqrt{d}$ and $\tilde{d}$ if the two values $\sqrt{d}$ and $\tilde{d}$ are far apart (i.e. by a power of $\log N$).


\subsection{Restriction to Sparse Graphs}
We are interested in sparse graphs -- i.e. those in which the degree of every vertex is $O(\polylog (N))$. If we use the random graph model above, and set $d = \Theta(\log^2 N)$, then if we allow all vertices to have an expected degree similar to $d$, then by Theorem \ref{theo:d_bounds}, $\|A\| = (1 + o(1))d$ almost surely, and the accuracies of both the classical and quantum algorithms are the same. Therefore, we are only going to see an advantage when we restrict the number of vertices that are allowed to have degree close to the maximum (which will be $O(\polylog N)$ by necessity). In general, in an effort to make $\|A\| = o(d)$, we should only allow at most $O(\log N)$ vertices to have degree close to the maximum, and the others must have asymptotically smaller (e.g. constant) degree. A class of graphs that satisfies this requirement is the class of power law graphs.

A distribution on power-law graphs is given in \cite{chung2003spectra} for which $d, \bar{d}$ and $\beta$ are parameters that can be varied freely. In graphs of this type, the number of vertices with degree $k$ is proportional to $k^{-\beta}$, and $d$ is the maximum expected degree of a vertex in the graph, while $\bar{d}$ is the average degree. We have the following results, also from \cite{chung2003spectra}:
\begin{enumerate}
\item For $\beta > 3$ and $d > \bar{d}^2\log^{3+\epsilon}N$, the largest eigenvalue of the graph is almost surely $(1+o(1))\sqrt{d}$, for some $\epsilon = O(1)$, and where $\bar{d}$ denotes the average degree.
\item For $2.5 < \beta < 3$ and $d > \bar{d}^{(\beta-2)/(\beta-2.5)}\log^{3/(\beta-2.5)}N$, the largest eigenvalue of the graph is almost surely $(1+o(1))\sqrt{d}$. 
\item For $2 < \beta < 2.5$ and $m > \log^{3/2.5 - \beta}N$, the largest eigenvalue is almost surely $(1 + o(1))\tilde{d}$.
\end{enumerate}
Note that in all of the above, the bounds still apply when the graph is sparse (i.e. $d = O(\polylog N)$). Hence, for power law graphs with exponent $\beta > 2.5$, we almost always get a quadratic improvement in accuracy over the classical algorithm. As the exponent decreases, so does the advantage gained by the quantum algorithm. 

Some interesting subclasses of power law graphs have exponents between 2 and 2.5. For example, `internet graphs' have exponents between 2.1 and 2.4, and the `Hollywood' graph has exponent $\approx 2.3$ \cite{faloutsos1999power}. In these cases, we might expect some quantum improvement over a classical approach, but not the full square root improvement.

\section{Acknowledgements}
CC was supported by the EPSRC. AM was supported by an EPSRC Early Career Fellowship (EP/L021005/1). No new data were created during this study.

\appendix
\section{Hamiltonian Simulation}\label{app:hamiltonians}
Here we give some more details of our (basic) approach to simulating a $\log$-local Hamiltonian $A$. 
\\
\\
Using the Lie-Trotter product formula, we have that, for any Hermitian matrices $H_1, ..., H_m$ satisfying $\|H_j\| \leq \zeta$ for all $j$,
\[
e^{iH_1}e^{iH_2}\cdots e^{iH_m} = e^{i(H_1+H_2+...+H_m)} + O(m^3\zeta^2),
\]
where the term $O(m^3\zeta^2)$ is used to denote some matrix $E$ such that $\|E\| = O(m^3\zeta^2)$. Applying this to the matrices $H_jt/p$ for arbitrary $t$ and some large integer $p$, we have
\[
\left|\left| e^{iH_1t/p}e^{iH_2t/p}\cdots e^{iH_mt/p} - e^{i(H_1+H_2+...+H_m)t/p} \right|\right| = O\left(m^3 \left(\frac{t\zeta}{p}\right)^2\right).
\]
Let $p \geq Cm^3t^2\zeta^2/\delta$ for some constant $C$. Then 
\[
\left|\left| e^{iH_1t/p}e^{iH_2t/p}\cdots e^{iH_mt/p} - e^{i(H_1+H_2+...+H_m)t/p} \right|\right| \leq \delta/p,
\]
and therefore
\[
\left|\left| \left(e^{iH_1t/p}e^{iH_2t/p}\cdots e^{iH_mt/p}\right)^p - e^{i(H_1+H_2+...+H_m)t} \right|\right| \leq \delta.
\]
Thus, to approximate $e^{iHt}$ up to accuracy $\delta$, it suffices to be able to implement the individual unitaries $e^{iH_jt/p}$ for $j \in [m]$, and $p = O(m^3t^2\zeta^2/\delta)$. If $\zeta \leq \|H\|$, and each individual unitary takes at most time $T$ to implement, then we can approximate $e^{iHt}$ up to accuracy $\delta$ in time $O(Tm^4\tau^2/\delta)$, where $\tau = t\|H\|$. 
\\
\\
An arbitrary unitary operation on $k$ qubits may be decomposed into a sequence of $O(k^2 2^{2k})$ one- and two-qubit gates \cite{nielsen2010quantum}. In order to implement such a unitary up to accuracy $\epsilon$ using some universal gate set, we must implement each individual gate up to an accuracy of $O(\epsilon/(k^2 2^{2k}))$, which, by the Solovay-Kitaev theorem \cite{dawson2006solovay}, can be achieved by using $O(\polylog((k^2 2^{2k})/\epsilon))$ gates from a universal gate set. Furthermore, the precise circuit implementing these unitaries can be computed classically in polynomial time \cite{dawson2006solovay}. Then the entire unitary may be implemented up to accuracy $\epsilon$ using a circuit of size $O(\poly(k,n,1/\epsilon))$.
\\
\\
In our case, the unitaries that we want to implement act non-trivially on $O(\log n)$ qubits. Since there are $m$ individual unitaries, and we apply each of them $p$ times, we must be able to implement each one to an accuracy $\epsilon = \delta / (mp)$ in order to implement the entire unitary $e^{iHt}$ up to accuracy $\delta$. 

By the above arguments, we can implement each unitary $e^{iH_jt/p}$ up to accuracy $\delta/(mp)$ in time that is polynomial in $m,p,n$ and $1/\delta$. Hence, we find that we can simulate $e^{iHt}$ up to accuracy $\delta$ in time 
\[
O(\poly(m,n,\tau, 1/\delta)).
\]

\section{Lipschitz Constants}\label{app:lipschitz}
Here we show that $f(x) = x^p$ is Lipschitz continuous over the interval $I=[-b,b]$, with Lipschitz constant $K = pb^{p-1}$. Since $f(x)$ is everywhere differentiable, it suffices to bound the absolute value of the derivative: $|f'(x)| = |px^{p-1}| \leq pb^{p-1}$ for all $x \in I$. 

To obtain a similar result for $g(x) = |x|^p$, we note that $g(x) = f(|x|)$. This is the composition of $f$ with the modulus function $h(x) = |x|$. The latter is Lipschitz continuous with Lipschitz constant 1, by the reverse triangle inequality: $\left| |x| - |y| \right| \leq \left| x - y\right|$ for all $x,y \in \mathbb{R}$. The composition of two Lipschitz continuous functions with Lipschitz constants $K, K'$ is also Lipschitz continuous with constant $KK'$. Hence, $g(x)$ is Lipschitz continuous over $I$ with Lipschitz constant $pb^{p-1}$. 
%
%

\bibliographystyle{abbrv}
\bibliography{../../../bibliography/references_new.bib}

\end{document}